\documentclass[dvipdfm,preprint,aps,amsmath]{revtex4}
\usepackage{graphicx}
\usepackage{bm}

\begin{document}
\preprint{}
\title{On radial geodesic forcing of zonal modes}
\author{A. Kendl}
\affiliation{Institute for Ion Physics and Applied Physics, Association Euratom-\"OAW, 
University of Innsbruck, A-6020 Innsbruck, Austria \vspace{2cm}}

\begin{abstract}
\vspace{1cm}
The elementary local and global influence of geodesic field line curvature on
radial dispersion of zonal modes in magnetised
plasmas is analysed with a primitive drift wave turbulence model. 
A net radial geodesic forcing of zonal flows and geodesic acoustic modes can
not be expected in any closed toroidal magnetic confinement configuration,
since the flux surface average of geodesic curvature identically vanishes.  
Radial motion of poloidally elongated zonal jets may occur in the presence of
geodesic acoustic mode activity. 
Phenomenologically a radial propagation of zonal modes shows some
characteristics of a classical analogon to second sound in quantum condensates.   
\end{abstract}

\maketitle

\section{Introduction}

Zonal modes play an important role in turbulent self organisation of 
quasi two-dimensional fluids, such as magnetised plasmas or in geophysical,
planetary and stellar fluid dynamics \cite{Diamond05,Fujisawa09,Manz09}. 
Zonal flows can be regarded as a low-frequency spectral condensate phase of
the turbulence \cite{Hallatschek00,Shats05} featuring the highest flow symmetry allowed by
the (generically inhomogeneous) system: atmospherical and oceanic zonal jets
are latitudinally extended flow structures, and zonal modes in
toroidal fusion plasmas appear as average flows on magnetic flux surfaces.  

Geophysical zonal flows can experience temporal variations and oscillations caused by
seasonal or topographical influences, and may migrate meridionally.
Similarly, plasma zonal flows in the inhomogeneous toroidal magnetic field are
modulated by finite frequency geodesic acoustic mode (GAM) oscillations
\cite{Winsor68}.  
The possibility of radial propagation of zonal flows and GAMs in fusion
plasmas has recently attracted interest, motivated by indicating experimental
observations \cite{Ido06,Lan08,Liu10}.

Previous studies on radial propagation included analytical theory of GAMs and zonal
flows \cite{Smolyakov00,Gao08,Miki10} and linear simulations of GAMs
\cite{Sasaki08,Xu08,Xu09}.  
Finite Larmor radius (FLR) effects in the presence of a background temperature
gradient have been identified as one potential mechanism for radial
propagation of zonal modes \cite{Itoh06,Zonca08,Sasaki08,Xu08,Xu09}.  
Further, it had been postulated that details of the magnetic field geometry,
in particular an up-down asymmetry, could cause radial GAM migration
\cite{Hager09,Hager10}.

In the following the particular influence of geodesic curvature and up-down
asymmetry on the radial dispersion of zonal modes is analysed.
A basic geometric property has to be recollected in this context: the flux
surface (zonal) average of geodesic curvature is zero in any toroidal configuration. 
Real and artificial geodesic forcing effects on zonal modes are discussed,
including the perils of using inadequately simple geometry models that violate
this property. 

The computational analysis shall here be based on a self-consistent nonlinear
model including both the drift wave turbulence fluctuation component and the
zonal mode condensate. Radial propagation can best be studied when radial
boundary conditions are eliminated by using a periodic domain. This however
excludes the utilization of 3-D flux-tube models, which rely on finite magnetic shear
to avoid unphysical flute modes but thereby introduce radial secularity. 

The most primitive model for self-sustained drift wave turbulence in
magnetised plasmas, that also allows radially periodic boundary conditions,
is the two-dimensional Hasegawa-Wakatani (HW) set of equations \cite{Hasegawa83}.  
This standard model is here extended to specifically study geodesic curvature
effects. 

The paper is organised as follows: 
in section II the curvature modifications to the HW model are introduced. 
The linear local radial dispersion relation of geodesically forced zonal HW
modes is discussed in section III.  
Numerical details regarding the nonlinear simulations are presented in
Chapter IV.
Computational results on local geodesic forcing of turbulence generated zonal
flows are reported in section V.
The essential flux-surface property of zero average geodesic curvature and
pitfalls concerning simplified geometry models are reviewed in section VI:
violation of this constraint results in unphysical radial propagation of zonal
flows and GAMs.
Poloidal variation of curvature is introduced in section VII, and effects on
apparent radial propagation of zonal modes is discussed in section VIII.
Finally, in section IX, a striking analogy to the phenomenon of ``second sound''
in quantum condensates is presented.

\section{Turbulence-flow model}

First, consider the HW model for drift wave turbulence \cite{Hasegawa83}.
It accounts for nonlinear instability driven by a gradient $\nabla n_0(x)$ in
plasma density and resistive parallel coupling between 
fluctuations of density $n$ and electrostatic potential $\phi$. 
The resulting turbulent state of $E \times B$ vortices in the $(x,y)$ drift plane
perpendicular to the magnetic field can form low-frequency $k_y=0$ zonal flow
structures with a finite wave number $k_x$.
The HW model is in the following employed with modifications for zonal flow
corrections on the dissipative coupling \cite{Dorland93} and including
magnetic field line curvature terms \cite{Scott05NJP}: 
\begin{eqnarray}
& \partial_t \Omega + [ \phi, \Omega ] & = d ( \hat \phi - \hat n )  
  - \kappa (n), \label{eq1} \\ 
& \partial_t n + [ \phi, n ] & = d ( \hat \phi - \hat n ) - g_n 
  \partial_y \phi + \kappa ( \phi - n ). \label{eq2} 
\end{eqnarray}
Standard drift normalisation is applied \cite{Scott05NJP}. 
The electrostatic potential $\phi$ is obtained from the vorticity $\Omega$ by
\begin{equation}
\nabla^2 \phi = \Omega. \label{eq.poisson}
\end{equation}

In the dissipative coupling term the zonal ($y$) average is subtracted from 
density and potential fluctuations
\cite{Dorland93}:
\begin{equation}
\hat \phi = \phi - \langle \phi \rangle, \qquad 
\hat n = n - \langle n \rangle.
\end{equation}
In toroidal plasma geometry the $x$ coordinate locally represents the minor radial,
and $y$ the poloidal direction. 

The original HW model (with $\kappa=0$) is determined by the dissipative coupling
parameter $d$ (proportional to $k_{||}^2$) and the density gradient scale
length $g_n$. The hydrodynamic limit of the Euler equations is recovered for
$d=0$ and $g_n=0$, while the adiabatic limit $d\gg1$ asymptotically
corresponds to the Hasegawa-Mima-Charney-Obukhov equation.
The general properties of the HW model have been extensively discussed
elsewhere (e.g. in  
Refs.~\cite{Koniges92,Biskamp94,Pedersen96,Zeiler96,Hu97,Camargo98,Korsholm03,Priego05,Numata07,Tynan07}).

The extended model here further includes field line curvature effects.
Normal and geodesic components of the magnetic curvature ${\cal K} = \nabla \ln
B$ enter the compressional effect on vortices due to magnetic field inhomogeneity by 
\begin{equation}
\kappa = \kappa_y  \partial_y + \kappa_x  \partial_x
\end{equation}
where the curvature components in toroidal geometry are
a function of the poloidal angle $\theta$. For a circular torus 
$\kappa_y \equiv c_B \cos(\theta)$ and $\kappa_x \equiv c_B \sin(\theta)$
when $\theta=0$ is defined at the outboard midplane. 
Pure interchange drive is acting for $\theta=0$ and pure geodesic effects
for $\theta=\pm \pi/2$, and mixed effects are achieved for intermediate angles. 
Slab turbulence is recovered for $c_B=0$. 

In the following, at first the local influence of geodesic curvature on zonal modes in
drift wave turbulence is discussed by choosing a constant angle $\theta$ for
the whole $y$-domain. Later, a periodic dependence of the form 
$\theta(y) = 2 \pi (y-L_y/2)/L_y$ will be introduced to globally map the
magnetic curvature inhomogeneities on a flux surface into the simulation domain.

\section{Linear analysis: local geodesic forcing of zonal modes}

Fourier analysis with an ansatz $n, \phi \sim \exp(i {\bf k} \cdot{\bf x} + i
  \omega t)$ and linearisation of eqs.~\eqref{eq1} and \eqref{eq2} for $k_y
  \ll k_x$
and $d=0$ (zonal modes) in the local approximation provides
\begin{eqnarray}
& \omega k_x^2 \phi_k & = \kappa_x k_x n_k  \label{fou1} \\ 
& \omega n_k & = \kappa_x k_x (\phi_k - n_k)  \label{fou2} 
\end{eqnarray}

For global simulations covering the whole flux surface (in an annulus or
flux-tube representation) the averages $\langle \kappa_x n_k\rangle_y$ and
$\langle \kappa_x \phi_k\rangle_y$ of the geodesic curvature operator acting
on the fluctuations would have to be taken into account in the linear analysis
(see section VII): sideband modes of the density then will result in the
well-known geodesic acoustic oscillations at the GAM frequency.  

In the present local discussion in this section, $\langle \kappa_x \rangle_y$
in itself is assumed to be nonzero and constant, and global sideband modes are neglected. 
This artificial scenario however can apply to zonal jet modes with nonzero but small
$k_y \ll k_x$ (rather than zonal flows with exactly $k_y=0$), and it corresponds to
artificial situations  where a finite flux-surface average geodesic curvature
results from an improper geometry model. 

The resulting local zonal wave equation has a frequency
\begin{equation}
\omega   = c_x k_x  ( \pi \pm \lambda )
\label{omega} 
\end{equation}
which is nonzero for finite average geodesic curvature, where $c_x = - (\kappa_x/2\pi)$
and $\lambda = (\pi/k_x) \sqrt{4+k_x^2}$.
The finite $k_x$ zonal modes are actually nonlinearly driven 
by turbulent Reynolds stress, which is eliminated in the zonal
dispersion relation eqs.~\eqref{fou1} and \eqref{fou2} by linearisation. 

The dispersion relation results in a phase velocity
\begin{equation}
v_{ph} = {\omega \over k_x} = c_x ( \pi \pm \lambda).
\label{vphex} 
\end{equation}
For $k_x \ll 1$ follows $\lambda \approx 2\pi/k_x$ so that $\omega
\approx 2\pi c_x$ is asymptotically independent of the wave number. 
Radial modulation of zonal modes is in this long wavelength limit obtained by
a local phase velocity  
\begin{equation}
v_{ph} \approx  c_x \lambda.
\label{vph} 
\end{equation}

The group velocity $v_{gr} = \partial \omega / \partial k_x \approx \pi c_x$
is for $k_x \ll 1$ much smaller than the phase velocity.
The following numerical simulations will show that zonal flows and GAMs mostly
appear with one dominant radial mode number propagating with $v_{ph}$. 
For single modes the concept of a wave packet travelling with a group velocity
is not relevant. Thus a Poynting approach on GAM propagation \cite{Hager09} is
pointless.  

In the following the simple analytical relation~\eqref{vph} is numerically
tested locally, and the effect of an artificial non-zero average geodesic
curvature $c_x$ on radial propagation of zonal modes is demonstrated and visualized.

\section{Nonlinear numerical solution}

Equations \eqref{eq1} and \eqref{eq2} are 
numerically solved in time with an explicit 3rd 
order Karniadakis scheme \cite{Karniadakis}, and the Poisson
brackets $[a,b] = (\partial_x a)(\partial_y b) - (\partial_y a)(\partial_x b)$ are
evaluated with the energy and enstrophy conserving Arakawa method
\cite{Arakawa}. The numerical method is equivalent to the one in
Refs.~\cite{Naulin03,Scott05NJP,Numata07}.  
Hyperviscuous operators $\nu^4\nabla^4$, with $\nu^4= -2\cdot10^{-4}$, are 
added for numerical stability to the right hand side of both equations
\eqref{eq1} and \eqref{eq2}, acting on $\Omega$ and $n$, respectively.

Eq.~\eqref{eq.poisson} is for a double periodic domain efficiently solved in
$k$ space by evaluation of $\phi_k = - \Omega_k / k_{\perp}^2$. 
Threaded FFTW3 libraries are employed for the 2-D 
forward and backward
Fourier transforms \cite{FFTW3}. 
The use of openMP multi-threaded FFTW3 routines on an 8-core workstation turned out to
be parallel efficient up to 4 threads only for the specified grid size
due to overhead costs. 

The equations are here discretised on a doubly periodic grid with various
(in general not quadratic) box dimensions. 
Spatial scales are in units of the drift scale
$\rho_s = \sqrt{T_e   M_i}/(eB)$ where $T_e$ is the electron temperature,
$M_i$ is the ion mass, and $B$ is the magnetic field strength. 
The gradient length scale $g_n= L_{\perp} / L_n \equiv 1$ is fixed by specifying
$L_{\perp} = L_n = |\nabla \ln n_o|^{-1}$. This normalises the
time scale to $t \rightarrow t \; c_s / L_n$. 
A stable time step for nominal parameters  on a $512 \times 512$ grid with
$L_x = L_y = 128 \rho_s$ scale was found to be $\Delta t = 0.0025$. 
The computations are initialized with a random bath of
quasi-turbulent density fluctuations, run into a saturated turbulent state
(typically obtained for $t>500$) and are continued for statistical analysis.

\section{Local results: radial propagation of zonal modes}

The radial zonal flow structure with respect to local normal and geodesic curvature
effects is in the following numerically analysed by varying the poloidal angle
$\theta$ through different simulations, while $\theta$ takes a constant value on the
respective ($x,y$) domains.

Nominal values for the dissipative coupling and curvature parameters used here
are $d=0.5$ and $c_B=0.05$.  
The flows are dominantly zonal for these parameters, and are blocked for lower
forcing ($d\lesssim 0.1$ and without interchange drive). 
The radial mode number $l$ of zonal flows increases with stronger forcing.

\begin{figure}
\includegraphics[width=10.0cm]{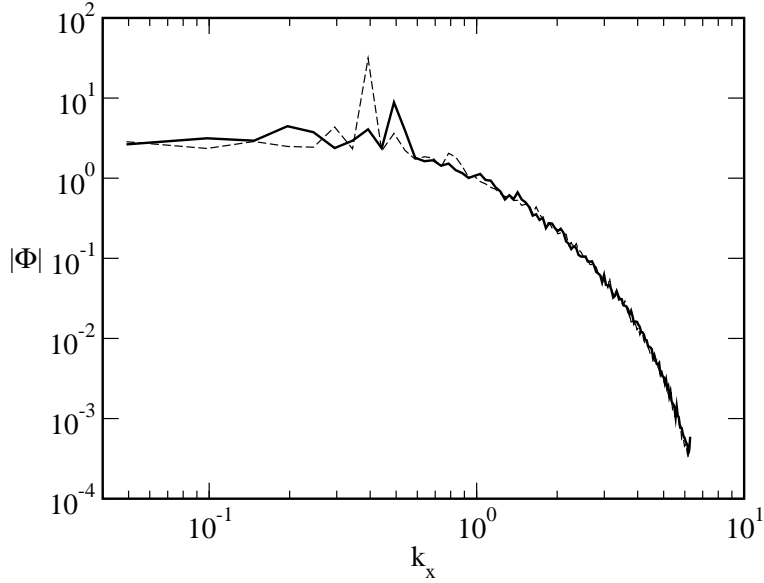}
\caption{\label{fig1} \sl Radial wave number $k_x$ (in units of the drift
  scale $\rho_s$) spectra of the potential $| \phi |^2$. Bold line:
mixed curvature effect ($c_B=0.05$, $\theta=\pi/8$);
thin dashed line: slab case ($c_B=0$).}
\end{figure}

Radial wave number spectra $| \phi | (k_x)$ for the $d=0.5$ flow dominated
scenario are shown in Fig.~\ref{fig1}:  
For the standard HW slab case ($c_B=0$, thin dashed line) the radial wave number features a
distinct peak at  $k_x \approx 0.39$ in units of $\rho_s$. This
corresponds to an $l=8$ radial mode of the ($m=0$, $n=0$) zonal flow with
radial wavelength $\lambda = 2 \pi / k_x \approx 16 \rho_s$.
Further contributions from $l=6$ and $l=10$ modes are an order of magnitude smaller.
The mixed curvature case ($\theta=\pi/8$, bold line) shows a strong maximum
for $l=10$ and a broader spectrum with similar amplitudes for $l=4$, 5, 7, and
8 modes.

\begin{figure}
(a)\includegraphics[width=9.0cm,draft=false]{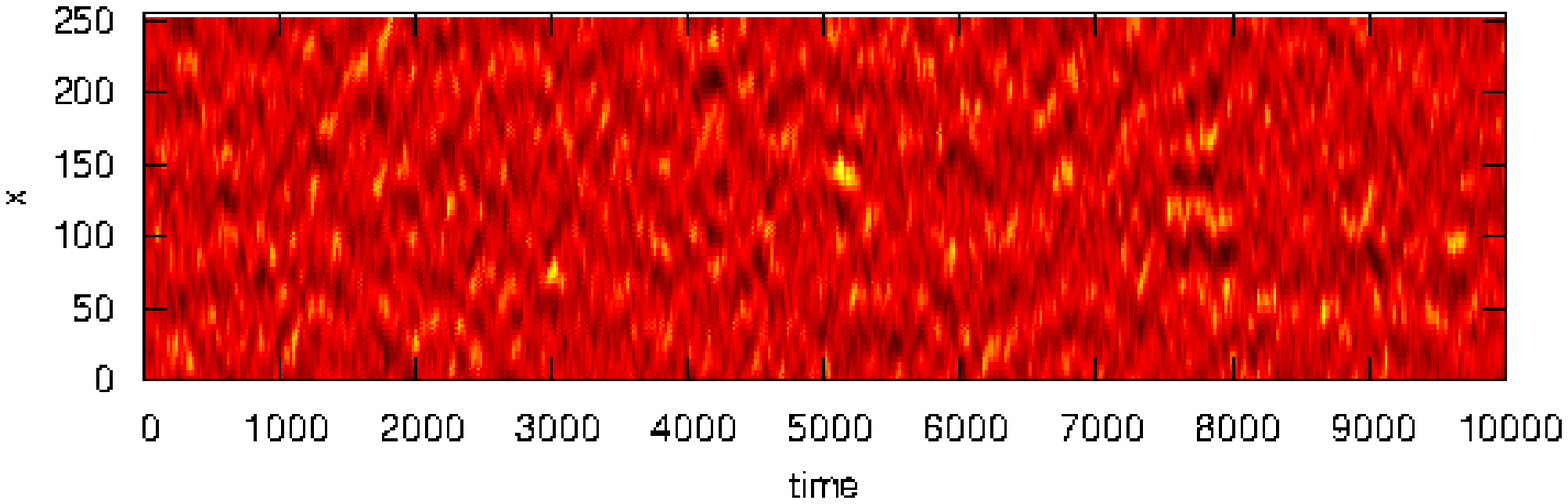}\\
(b)\includegraphics[width=9.0cm,draft=false]{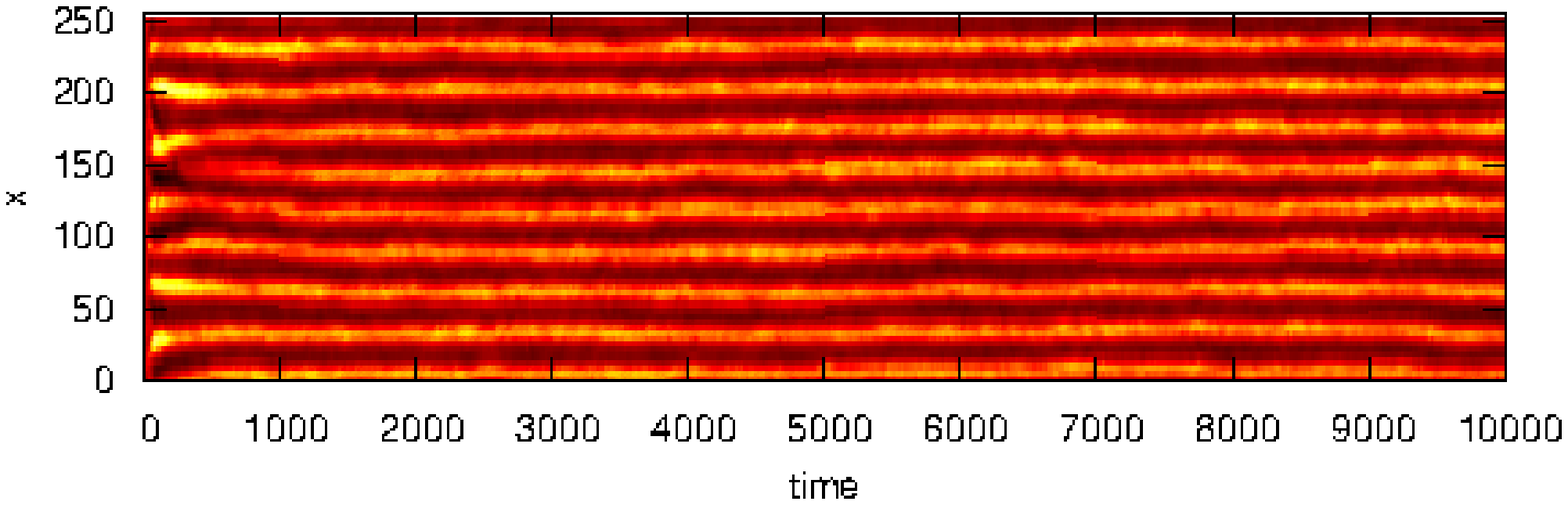}\\
(c)\includegraphics[width=9.0cm,draft=false]{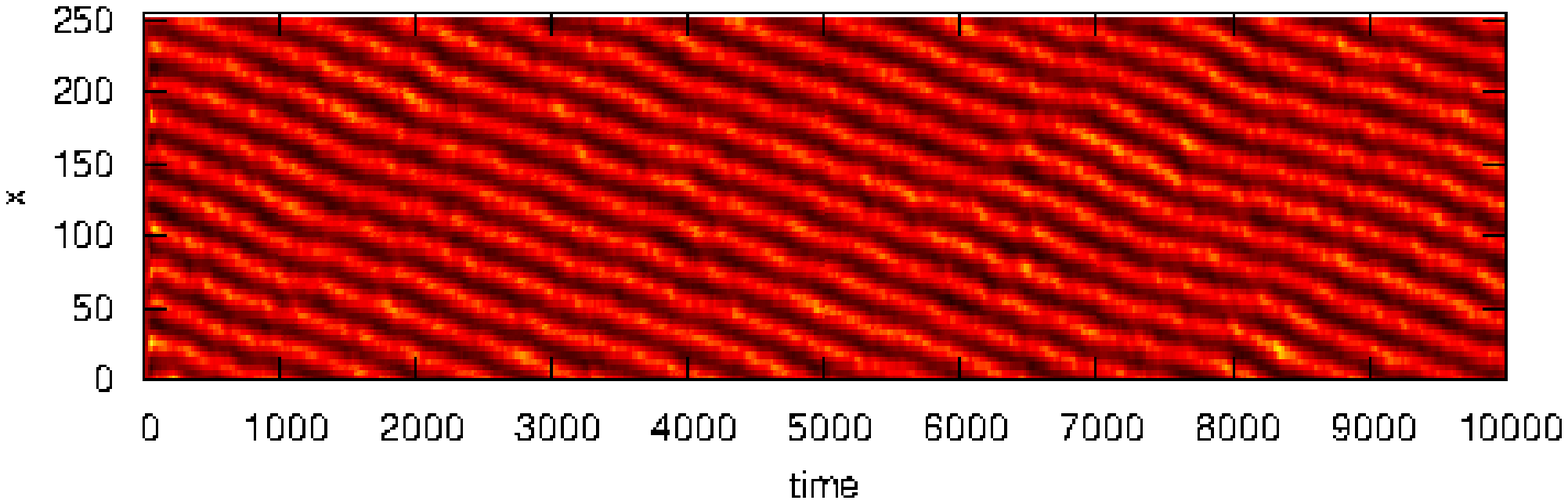}\\
(d)\includegraphics[width=9.0cm,draft=false]{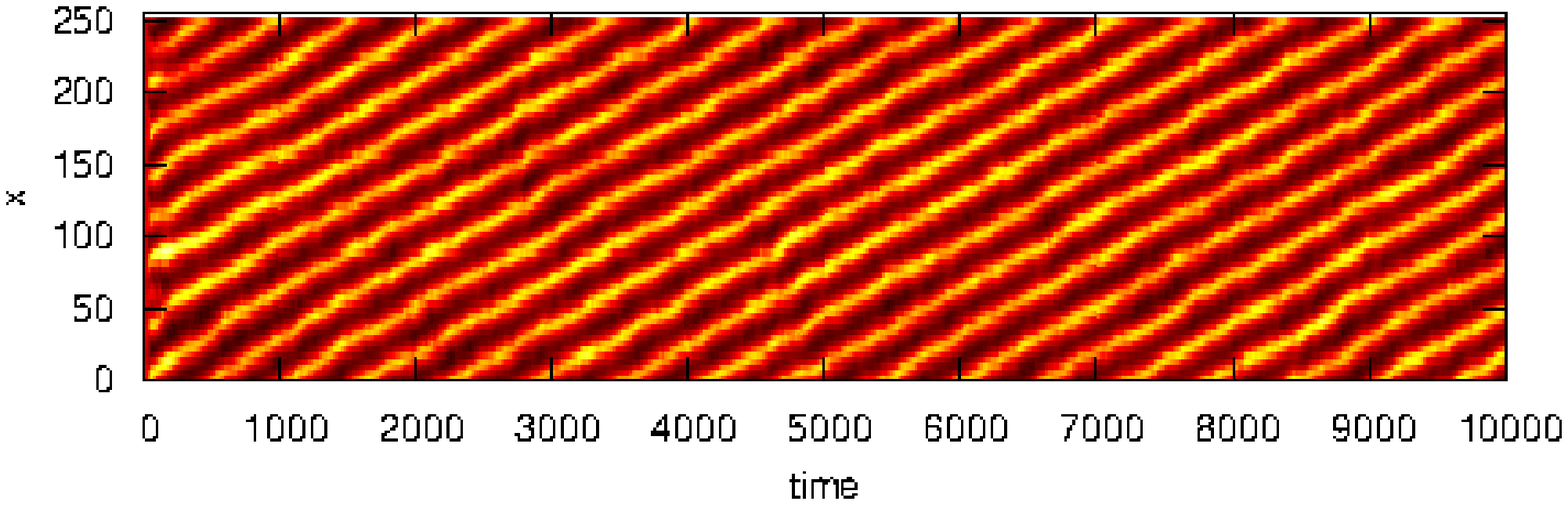}\\
(e)\includegraphics[width=9.0cm,draft=false]{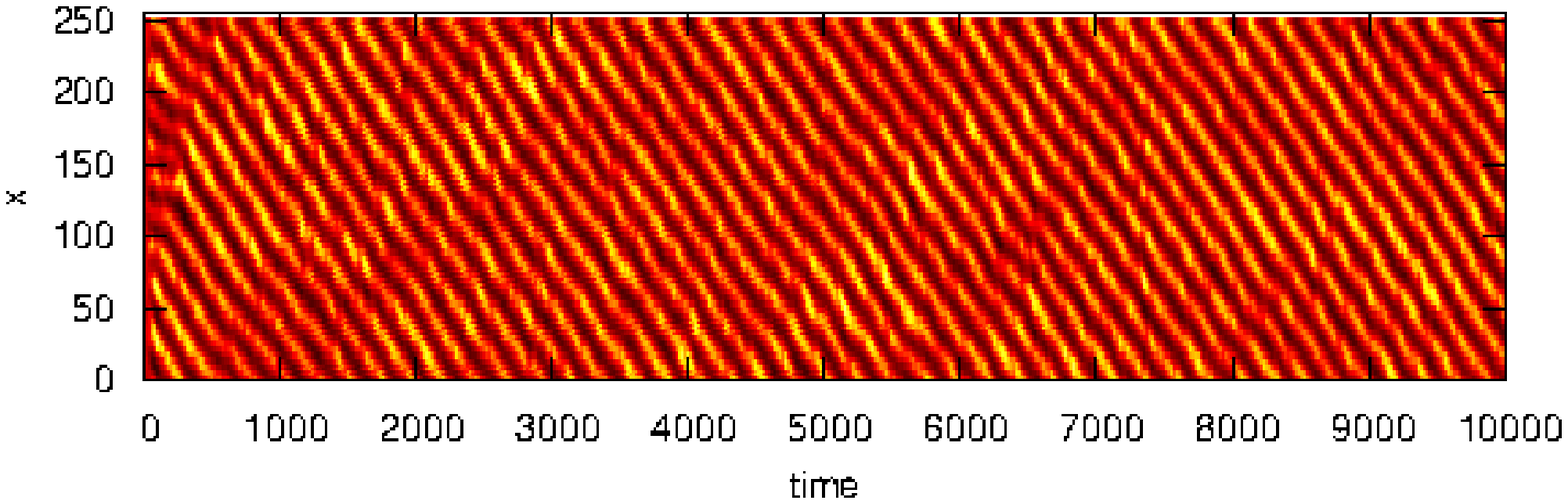}
\caption{\label{fig2} \sl Zonal potential $\langle \phi \rangle (t,x)$ for various
  cases of forcing and curvature parameters: \\
  (a) blocked flow for low forcing ($d=0.1$, $c_B=0$); 
  (b) zonal flow for high forcing ($d=0.5$, $c_B=0.05$, $\theta=0$);  
  (c) mixed curvature case ($d=0.5$, $c_B=0.05$, $\theta=\pi/8$); 
  (d) same but reversed field ($d=0.5$, $c_B=0.05$, $\theta=-\pi/8$);
  (e) pure geodesic drive ($d=0.5$, $c_B=0.05$, $\theta=\pi/2$).}
\end{figure}

This radial structure of the zonal flow modes is also clearly visible in 2-D
($t$,$x$) plots for cases with strong forcing. 
Fig.~\ref{fig2} shows the time evolution of the zonally averaged potential for
various drive parameters and poloidal angles. 
Here the spatial resolution is reduced to 256 $\times$ 256 grid points for
the same physical domain size of $L_y = 128 \rho_s$ as before. Runs are now
taken to $t=10^4$.

As predicted above, the zonal modes propagate radially with a velocity
proportional to the geodesic curvature parameter $c_x$ and to the wavelength
$\lambda = L_x/l$. 
The phase velocity measured in the nonlinear simulation is in very
good agreement with the estimate obtained from linear theory.
For case (e) with $c_x = - c_B / (2\pi) \approx - 0.008$ and 
$\lambda/\rho_s = 128/8 = 16$ 
we  expect $v_{ph} \approx  c_x \lambda \approx 0.064$ from equation~\eqref{vph}, 
while the simulation gives $v_{sim} = \Delta x / \Delta t = 128 / 1860 \approx 0.069$.

At this point it should again be stressed that this local result as a test for
eq.~\eqref{vph} does not apply to actual toroidal $k_y=0$ zonal flows. 
It will however be relevant for the following discussion on geometrical
artefacts and on the interpretation of spatio-temporal features identified as
radially propagating zonal jet structures.

\section{Flux-surface property of geodesic curvature}

In a torus the ``upper'' and ``lower'' regions locally correspond
to positive and negative $c_x$, respectively, if the poloidal coordinate
$\theta$ is chosen as here to be aligned ``upwards''. 
Reversal of the magnetic field direction will locally also
reverse the sign of the geodesic term through $c_x (-\theta) = - c_x(\theta)$.
For a circular torus $\kappa_x = c_B \sin(\theta)$. 

The poloidally local effects of radial geodesic forcing
globally balance, if positive and negative contributions from geodesic
curvature cancel by integration over a flux surface. 
This is indeed the case for any low beta toroidal MHD equilibrium, including
up-down asymmetric tokamaks or 3-D stellarator configurations.
The flux surface average of the geodesic curvature always identically
vanishes \cite{Hazeltine85}. 

\begin{figure}
(a) \includegraphics[width=9.0cm]{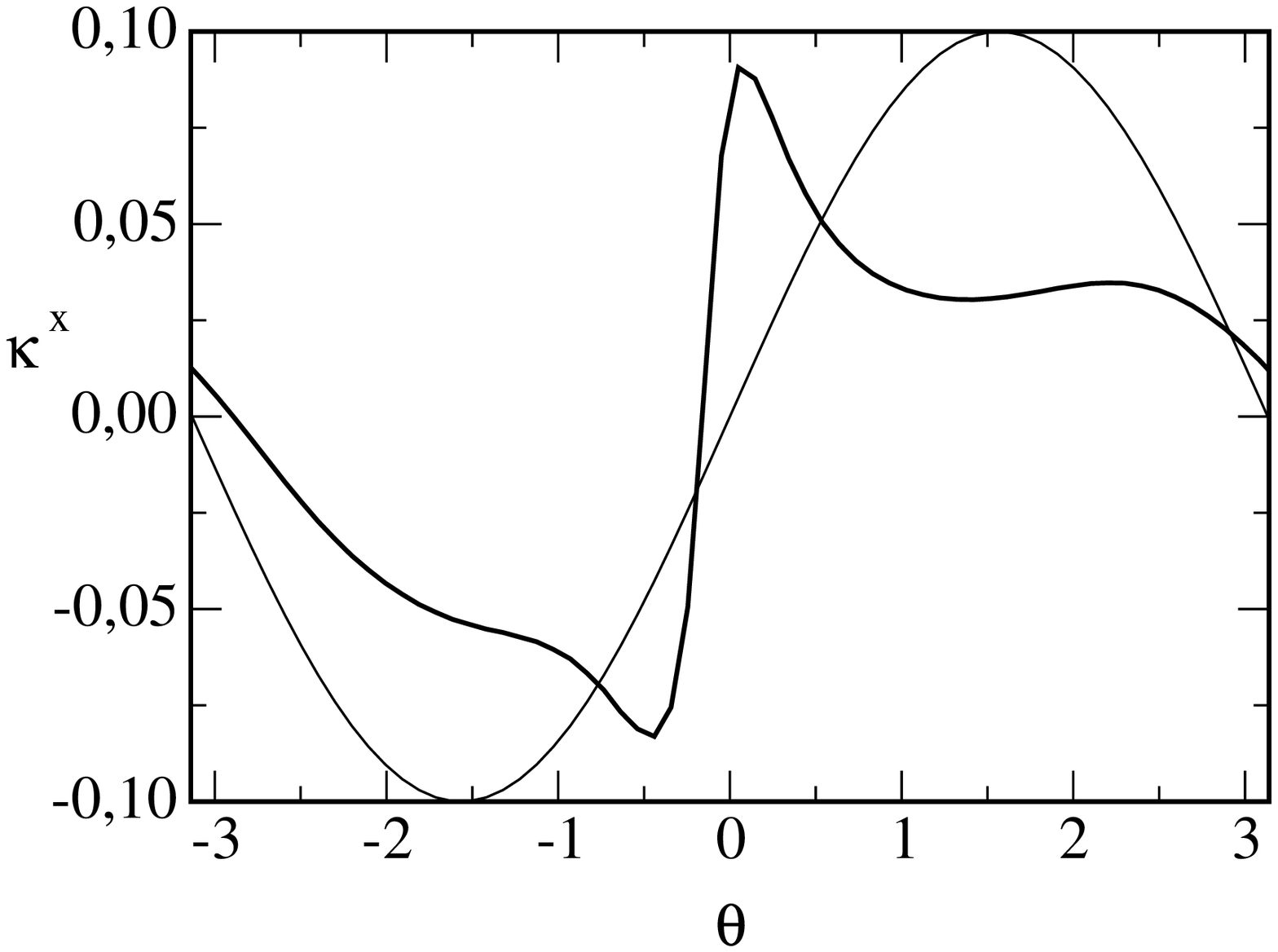}\\
(b) \includegraphics[width=9.0cm]{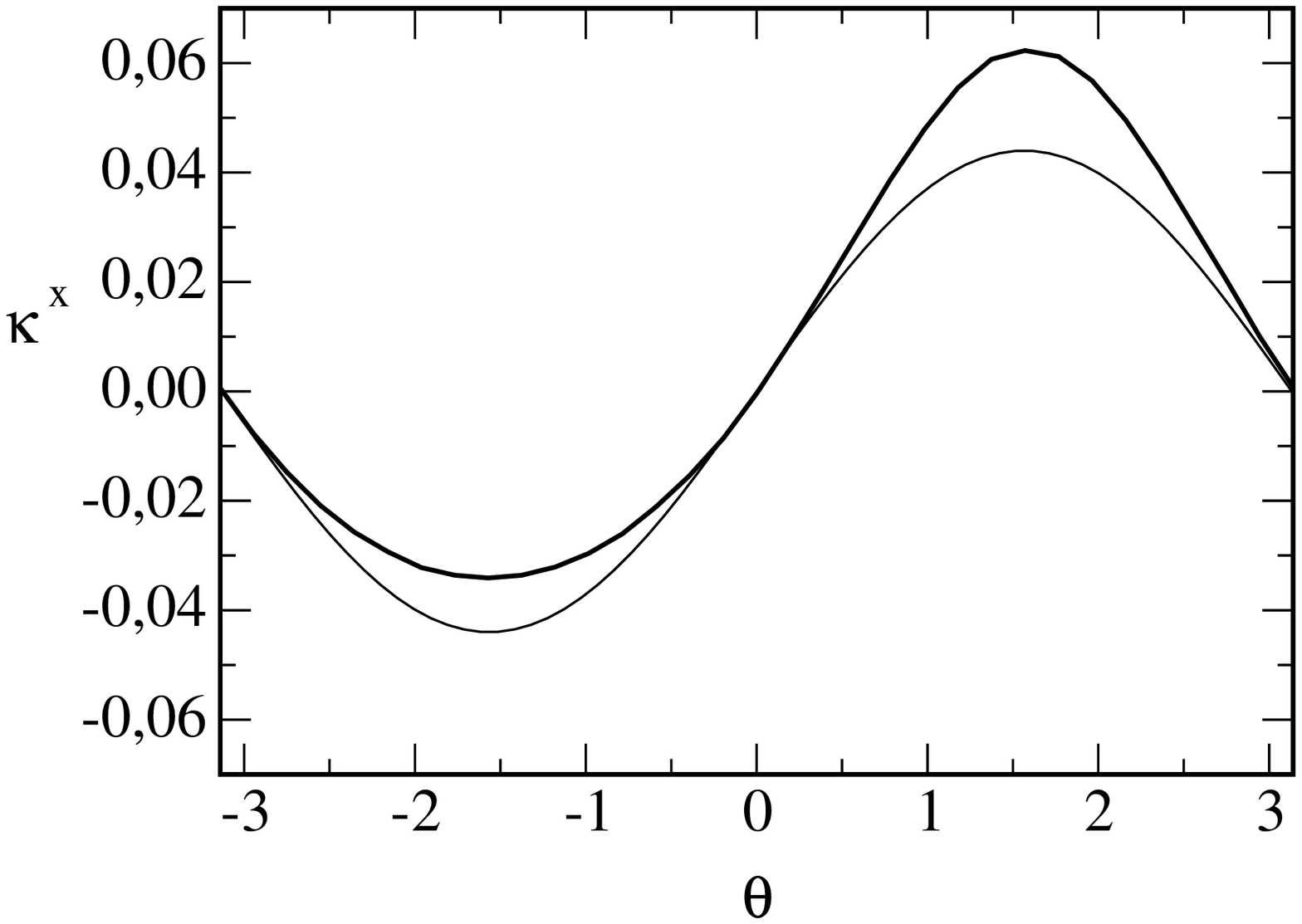}
\caption{\label{fig4} \sl
(a) Geodesic curvature $\kappa_x (\theta)$ in a circular torus (sinusoidal thin
curve) and in a realistic ASDEX Upgrade equilibrium (bold curve). The average
always is identically zero.\\
(b) Geodesic curvature $\kappa_x (\theta)$ resulting from an inappropriately simple X-point
geometry model (bold curve), which leads to an unphysical finite average.}
\end{figure}

This is illustrated in Fig.~\ref{fig4} (a) where the geodesic curvature $\kappa_x
(\theta)$ as a function of the poloidal angle is shown for the example of a real up-down
asymmetric ASDEX Upgrade equilibrium in a lower single-null configuration,
calculated with the VMEC equilibrium code \cite{Strumberger} at a radial position
$\rho=V/V_o=0.95$ in volume coordinates. 
The maximum of geodesic curvature is reduced in the vicinity of the X-point
location $\theta=+\pi/2$, but the average over $\theta$ remains zero as expected.
This peak reduction affects the GAM and zonal flow amplitudes (and
consequently turbulent transport) through the same mechanisms as by the
up-down symmetric effect of flux-surface elongation \cite{Kendl03,Kendl05,Kendl06}. 

In no toroidal configuration a net radial propagation of ($m=0$, $n=0$) zonal
modes by geodesic forcing alone is therefore possible. 

Approximate analytical toroidal geometry models (e.g. when introducing
up-down asymmetry effects) have to account for this cancellation, otherwise
unphysical radial propagation may appear in simulations. 
This is illustrated in Fig.~\ref{fig4} (b): here $\kappa_x (\theta)$ (bold line) has been
calculated from the simple up-down asymmetric model geometry from
Ref.~\cite{Hager10}, where flux surfaces are vertically shifted. It can be
clearly realised that this corresponds to a shift $\kappa_x (\theta) \approx c_0
+ c_B \sin \theta$ with the result of a nonzero average geodesic curvature. 
This unphysical finite average acts on GAMs and zonal flows exactly like the
local model discussed in Sections III and IV and results in an artifical
radial propagation. Other simplified analytical geometry models in general
also do not conserve the zero average geodesic curvature property. 

Most results concerning the overall turbulence and flow levels and other
characteristics from nonlinear simulations are usually not or only to a very small
degree affected by this particular inconsistency. 
Simple X-point models for example well allow to study an influence of local
magnetic shearing on turbulence and transport. 
For any discussion of radial propagation of GAMs and zonal flows
the use of analytical model geometries has to either explicitly account for
a zero average or is otherwise inappropriate. Then numerical solutions from
codes that solve actual MHD equilibria are required.

\section{2-D geodesic acoustic modes} 

Next, a poloidal variation of the geodesic curvature within the 2-D HW
computational domain is taken into account. The extension $L_y$ is elongated up
to $L_y/L_x = 16$ in order to cover a complete poloidal circumference.
The curvature terms $\kappa_y \equiv c_B \cos(\theta)$ and $\kappa_x \equiv
c_B \sin(\theta)$ are now periodically varying along the $y$ coordinate by
$\theta(y) = 2 \pi (y-L_y/2)/L_y$. The average along $y$ can so be interpreted
as the flux surface average, while still any parallel (3-D) dynamics is neglected.

The linear spectral zonal HW equations now become
\begin{eqnarray}
& \omega k_x^2 \langle \phi_k \rangle & = \langle \kappa_x k_x n_k \rangle \\ 
& \omega \langle n_k \rangle & = \langle \kappa_x k_x (\phi_k - n_k) \rangle  
\end{eqnarray}
where the brackets denote the average along $y$. 
Using the identity $\sin^2 \theta = 1- \cos^2\theta = {1\over 2} [1- \cos
  (2\theta)]$ and defining the zonal flow velocity by $u_k \equiv i k_x
\langle \phi_k \rangle$, the dispersion relation gives
$\omega^2 u_k = {1 \over 2} c_B^2 u_k$.
The result is a periodic oscillation of the zonal flow with the GAM frequency
$\omega = c_B / \sqrt{2} \equiv \omega_{GAM}$. 

The present 2-D HW model lacks any parallel effects of field line connection and
wave dynamics. In particular, the absence of coupling from the density
sideband to parallel dissipative sound and Alfv\'en dynamics overestimates the
GAM amplitude, compared to 3-D flux-tube simulations \cite{Scott02}. 
On the other hand, the density sideband is also in 2-D nonlinearly coupled to the
density cascade and its respective small-scale dissipation. 
The geodesic transfer thus constitutes a sink mechanism for zonal flow energy,
and the 2-D toroidal zonal flow amplitude is less than in the slab or local
case (and also less than in 3-D toroidal simulations). 

The results are clearly quantitatively different from more complete 3-D
(drift-Alfv\'en or gyrofluid) models, but the emphasis here is on principle
understanding of basic curvature coupling mechanisms which do not specifically
rely on parallel dynamics. 

\section{Spatiotemporal zonal flow pattern: apparent radial motion} 

\begin{figure}
\includegraphics[width=12.0cm]{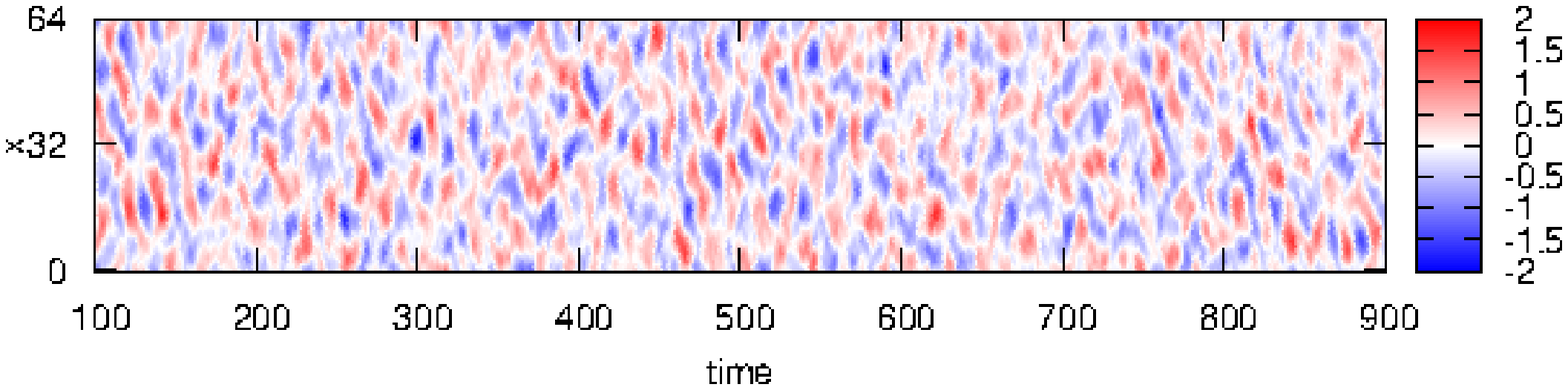}\\
\includegraphics[width=12.0cm]{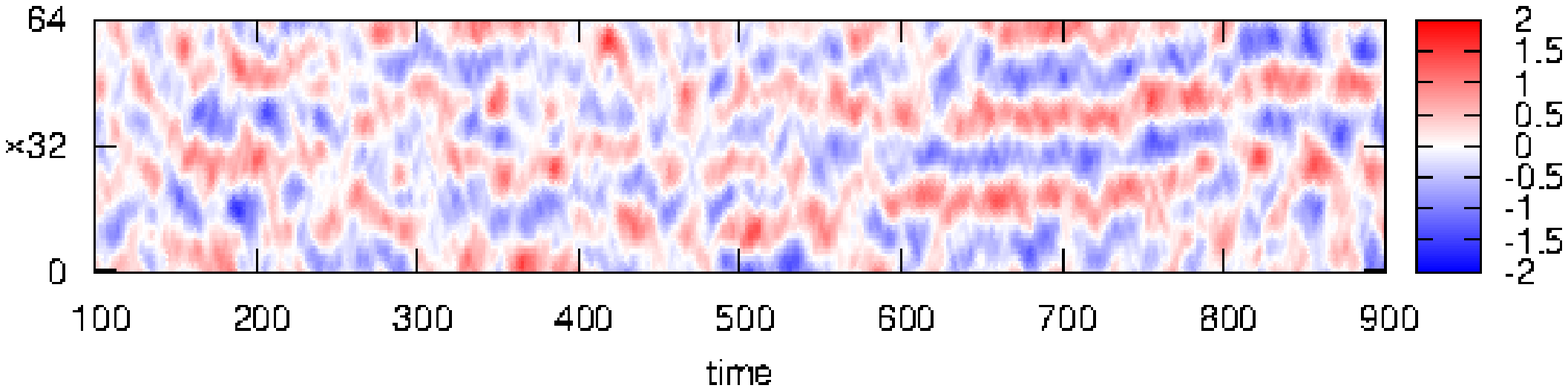}
\caption{\label{fig-hc3} \sl Pattern of zonal flow velocity $\langle
  v_y\rangle (x,t)$ for $c_B=0.5$ (top) and  $c_B=0.25$ (bottom). 
Intermittent diagonal streaks bear semblance of radial ``propagation'' for
high-frequency GAMs at $c_B=0.5$ but are absent for lower frequencies.} 
\end{figure}

The poloidally extended model for $\kappa = \kappa(y)$ is now solved numerically
with simulation parameters $d=0.5$, $g_n=1$, $n_x=128$, $n_y=2048$, $L_x=32\rho_s$,
$L_y=512 \rho_s$, $\nu^4 = -5\cdot 10^{-4}$, $\Delta t = 10^{-4}$, and with a
curvature parameter $c_B=0.5$. The zonal flow velocity $\langle
v_y\rangle (x,t)$ is shown during the initial saturated phase of the
computation ($100<t<900$) in Fig.~\ref{fig-hc3}. 

The plot for $c_B=0.5$ features an irregular checkerboard like pattern: quasi
periodic oscillations appear both in space and time as the GAM modulation
reverses the zonal flow. The radial wave number of the pattern is in the order
$k_x \sim 1$ and the frequency in the order of $\omega \sim \omega_{GAM}$. 
When the curvature parameter is reduced to $c_B=0.25$ (with otherwise
identical simulation parameters) the frequency scales proportionally but the
modulation appears weaker and the zonal flow direction at a specific radial
location $x$ rarely reverses with time. 

Another feature of the pattern for $c_B=0.5$ immediately strikes the eye: 
some zonal flow structures appear connected diagonally and give the semblance
of moving radially in time with a velocity $v_x^{app} = \omega / k_x$. 
The connections first appear randomly with no prefered radial direction. 
This in-out symmetry can be spontaneously broken and an effective diagonal
pattern may emerge transiently (as can be seen for around $t \approx 100$ and
$t \approx 500$).  Occurrence and directionality of such symmetry
breaking appear to be random and to depend on simulation parameters and resolution. 

The resulting apparent radial motion however is illusionary as the actual
spatio-temporal zonal checkerboard-like structures do not move. 
This appearence may mislead to the conclusion of GAMs propagating radially
with  $v_x^{app} \sim \omega_{GAM} / k_x$. 
But as it turns out it is not the GAM modulated zonal flows (or the
GAMs) which are moving, but rather $k_y \ll1 $ vortices.

\begin{figure}
\includegraphics[angle=270,width=5.0cm]{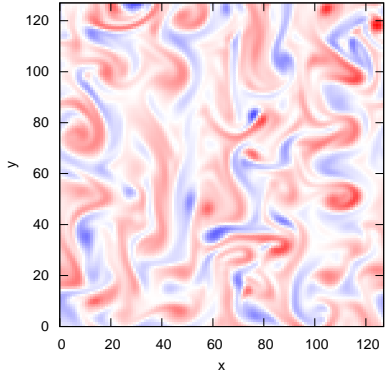}
\caption{\label{fig-jets} \sl 
``Zonal jets'': $k_y\ll1$ vortices strained out by zonal flow shearing (in
  particular around $x \approx 50$). 
  Vorticity $\Omega(x,y)$ is shown (negative values: blue; positive
  values: red; in arbitrary units) on a part of the computational domain.} 
\end{figure}

Vortex structures can be detected in a wide spatial range, from circular
eddies in the size of $k_y \sim k_x \sim 1$ to $k_y=0$ zonal flows. In an
intermediate range poloidally elongated vortices with $k_y \ll1 $ are
visible. These ``zonal jet'' structures appear strained out by the zonal flow
shear, like shown in Fig.~\ref{fig-jets}. 

Zonal jets are, in contrast to $k_y=0$ zonal flows, susceptible to geodesic
forcing. The poloidally elongated structures can experience local forcing in
the radial direction.  
Diagonal connections in the $\langle v_y\rangle (x,t)$ spatiotemporal GAM
pattern appear through local radial zonal jet migration if the life time of the jets
is comparable to the GAM period. For lower $c_B$ (e.g. 0.1) the vortex
turnover period is much smaller than the GAM modulation time of the zonal
flows, and no diagonal motion is observed in spatiotemporal diagrams. 

Up-down asymmetry will not change the average effect of geodesic forcing on
($k_y\equiv0$) zonal flows and GAMs. 
It may however have an indirect effect on poloidally localized ($k_y \ll 1$) zonal jets. 
Asymmetric shaping also influences local magnetic shear, which is, for
example, strongly enhanced near a (lower) single-null X-point \cite{Kendl03}.  
Zonal jets emerging through straining of drift wave vortices can be weaker at
the (downward) poloidal location of strong magnetic shear, 
and may experience less geodesic forcing than on the other (up) side. 
Under such circumstances an effective geodesic forcing of zonal jets might
actually appear. In order to consistently account for local magnetic shear
effects in interplay with zonal flow shearing a 3-D (flux-tube) model is
indispensible. 
This case is therefore not further considered within the simplified 2-D model
of the present discussion. 

It has been demonstrated in the preceeding sections that unphysical radial
propagation of zonal flows and GAMs by geodesic forcing may apparently occur
for certain computational set-ups. 
Radial propagation of zonal modes by other possible causes (for example by
inhomogeneous FLR effects) is not disputed here.

\section{``Second sound'' in drift wave turbulence}

Radially propagating zonal modes, however they may originate, bear a striking
resemblence to the phenomenon of ``second sound'' in quantum codensates. 
Second sound in two-fluid systems, for example in superfluid He, describes the
transport of temperature and entropy in the form of waves rather than by
diffusion \cite{Enz74,Joseph89}.  
Phonons (``first sound'') are modulated by temperature waves (``second
sound''), while the normal component and the condensate oscillate out of phase. 

Drift wave turbulence can be considered within the framework of 
wave kinetics \cite{Diamond-Book}. 
Wave density and temperature in this context do not refer to
the fluid moments of the plasma particles, but rather to the population
density and wave energy of the drift mode excitations. 

The spectral drift wave mode density of the turbulence at a specific radial
location is modulated in time by the passing wave of the radially propagating
zonal flow condensate.  
Although the condensate phase does not by itself participate in radial plasma
transport this leads to a radial wave like transfer of spectral mode energy,
when the turbulent phase is co-existing with the propagating condensate.  
Phenomenologically this propagation shows characteristics of a classical analogon 
to second sound in quantum condensates: in a wave kinetic picture a mode
temperature of the turbulent phase may be defined by the mean spectral energy,
which is modulated by zonal condensate mode propagation in the form of a radial wave
rather than by diffusive spreading.  

The source of drift wave turbulence is, with respect to its two-dimensional
character and dual cascade property, located within the spectral
distribution around $k_{\perp} \rho_s \sim 1$ and is for many cases well
separated from both dissipation and modulation (condensate) scales. 
Weakly damped second sound in drift wave turbulence of this form appears
possible, in contrast to general 3-D turbulent Kolmogorov
type systems \cite{Falkovich90}.
A future more rigorous theoretical consideration of this phenomenon surely is eligible.

\section{Conclusions}

Basic mechanisms of local and global radial geodesic forcing on zonal modes have
been discussed with the 2-D HW standard model of drift wave turbulence.
The main results are summarized as follows:

\begin{itemize}
\item Radial propagation of $k_y\ll1$ zonal modes can occur by local geodesic
  forcing, which has been demonstrated by poloidally local simulations.
\item A net radial geodesic forcing of $k_y=0$ zonal modes (zonal flows and
  GAMs) can not be expected in any closed toroidal magnetic confinement
  configuration, as the flux surface average of geodesic curvature always
  identically vanishes. 
\item Unphysical radial propagation of zonal flows and GAMs may appear in
  simulations that use inappropriately simple geometry models violating this zero
  average property.
\item Apparent radial motion of spatio-temporal zonal mode patterns may be visually
  diagnosed in the presence of geodesic acoustic oscillations. 
\item Zonal jets with $k_y\ll1$ occur as vortices strained out by $k_y=0$ zonal
  flows or GAMs and may experience local geodesic forcing.
\item Radially migrating poloidally localized zonal jets lead to
  a diagonal connection in spatio-temporal GAM patterns, and in conclusion to
  an illusion of radial GAM propagation.
\item Up-down asymmetry does not alter the average geodesic curvature from
  zero, and therefore does not affect radial geodesic forcing of zonal flows
  and GAMs.
\item Up-down asymmetry may influence the local amplitude of zonal jets
  through local magnetic shearing. 
\item Phenomenologically a radial propagation of zonal modes shows some
characteristics of a classical analogon to second sound in quantum condensates. 
\end{itemize}

While the applicability of the 2-D HW model on realistic fusion
plasmas is definitely limited, it has served to elucidate basic physical
effects of the magnetic field geometry on drift wave turbulence and zonal flows. 
Any future computational studies with more complete models on radial
propagation of zonal modes should not rely on inappropriately simplified
geometry models but have to make use of actual MHD equilibria.


\footnotesize
\section*{Acknowledgements}

The author thanks B.D.~Scott (IPP Garching) for valuable discussions and
S. Konzett (University of Innsbruck) for model geometry calculations.
This work was partly supported by the Austrian Science Fund (FWF) project
Y398, by a junior research grant from University of
Innsbruck, and by the European Communities under the Contract of
Associations between Euratom and the Austrian Academy of Sciences, 
carried out within the framework of the European Fusion Development
Agreement. The views and opinions herein do not necessarily reflect those of
the European Commission.


\end{document}